# Recent advances in triboelectric nanogenerators: energy harvest and other applications

L.J. Zhang

Triboelectric nanogenerators (TENG) with triboelectrification and electrostatic induction effects have attracted wide attention in recent decades, for its diversified application scenarios such as power generation, sensing, and so on. Undoubtedly, although still lacks standardized large-scale production, TENG has demonstrated good performance and increasingly promising prospects in fields such as energy harvest, healthcare, and medical monitoring. Here, this minireview provides a brief summary on the latest application progress of TENG, which may offer insights for expanding application fields and developing design concepts for TENG based devices.

## 1. Energy Harvesting

Carbon emission reduction and carbon neutrality have become the common goals worldwide in facing with the threat of environmental change and fossil energy depletion. With the continuous advancement of the development and utilization of clean energy, new advances in the use of triboelectric nanogenerators (TENG) for capturing clean energy have also been reported. The triboelectric nanogenerators network with wheel-structure has been demonstrated to be an efficient tool for low-frequency and small-amplitude ocean wave energy harvest.[1] With blades rolling on the water surface like a wheel and hyperelastic networking structure stretching and shrinking like a spring, TENGs can be agitated and form a large-scale network. In addition, a three-dimensional chiral network of TENGs for water wave energy harvest was proposed this year. with chiral connections between unbalanced units and wave-absorption behavior, the network is endowed with scalability and energy absorption performance from various directions and depths of water.[2]

Moreover, liquid-solid TENGs with dynamic electric-double-layers was also reported to be stable water wave energy harvesters, presenting output current, voltage, and average power density of 60 μA, 60 V, 5.38 W m$^{-3}$, respectively.[3] Furthermore, a unidirectional rotation cylindrical TENG with output current of 30 μA and output power of 50 mW was also reported recently, which is capable of powering various devices including LEDs and portable anemometer et al.[4]

In addition to wave energy, wind energy, including highway wind energy generated by highway moving vehicles, is also an important source of renewable clean energy with large reserves. Wang's group proposed a TENG based energy harvester, achieved satisfactory performance with power density of 0.2 W m$^{-2}$ (wind speed 3 m s$^{-1}$) via single TENG module.[5] It is worth noting that the TENG harvester exhibits good stability with no significant performance degradation in 1440000 operations, paving the way for its potential application in the energy supply field for environmental sensors and other devices.

Gentle wind energy acts as a valuable component in clean energy, but the low speed makes its harvest process via conventional wind power generation equipment an enormous challenge. Recently, a leaf-like TENG(LL-TENG) was developed to resolve this problem, employing contact electrification resulting from the damped forced vibration. With a unit of only 40 cm$^2$ and a low resonant frequency of 4 Hz (wind speed 2.5m s$^{-1}$), LL-TENG presents satisfactory open-circuit voltage (338 V), short-circuit current (7.9 μA), transferred charge density (62.5 μC m$^{-2}$) and peak power (2 mW).[6]

Biomechanical energy such as hand motion has also been reported to be harvested by TENGs. Array of paralleled TENGs exhibits enhanced electrification efficiency and reliability and higher output than those serially connected, providing promising methods for electronic devices power supply.[7]

## 2. Cancer Therapy and medical monitoring

Recently, an implantable, biodegradable, and wireless electric device based on TENGs was developed for cancer treatment.[8] Ultrasound triggers electric field generation and drug release performance of the TENGs in a wireless way, resulting in cell mitosis arrest and cell death enhancement. Moreover, TENGs can be degraded with the excitation of ultrasound, thus avoids the secondary surgical extraction activity. In addition, a drug-free tumor therapy based on fabric TENGs, Triboelectric Immunotherapy, has also been developed recently. By employing pulsed direct current generated by the triboelectrification and electrostatic breakdown effects of TENGs, 4T1 solid tumor inhibition can be realized through 4T1 cells death, T cells-mediated immunity response activation, cytokines accumulation around tumor, indicating the promising anti-tumor effects of this portable and wearable system.[9]

Contactless medical monitoring system based on TENGs, made from biodegradable lignocellulosic, was also developed by Wang's group recently. The TENGs exhibit low cost, environmental friendliness, and good recyclability, profiting from abundant and accessible raw materials.[10]

Head impacts can cause concussions with cognitive, affective functional impairment. Based on TENGs, a real-time head impact monitoring device with multiangle self-powered sensor array was recently developed. The TENGs monitoring device presents satisfactory performance, not only including impact sensing with resolution of 1.415 kilopascals, response time of 30 milliseconds in the range of 0-200 kilopascals, but also offering reconstructed mapping and injury grade of head impact via impact force to electrical signals conversion.[11]

Biometric gait monitoring and recognition provides a significant tool for assessment for muscular and neurological function, since gait can be easily influenced by multiple factors such as brain cognition and function, balance ability, etc. It's reported recently that a multi-point body motion sensing network based on TENGs and textile fabrics demonstrated high accuracy for recognition of five pathological gaits with the assistance of machine learning.[12] Rehabilitation exercise activity on auxiliary exercise system can also be monitored with the sensing network, thus recovery training extent and instruction information can be provided.

Cost-effective and fast fluid pressure measure through portable devices is a favorable method that have attracted considerable attention. A fluid pressure sensor based on a liquid piston TENG (LP-TENG) was proposed recently, realizing an accuracy of 0.4 kPa (in the range of 0-30 kPa) and resolution ratio of 10 mm Hg, which is capable for blood pressure measure in healthcare field. Potential applications can also be expected in civil service because the performance of LP-TENG in fluid state measure, such as direction and rate.[13]

Respiratory signals, one of the important physiological information, have been used for Health monitoring and disease prediction. A TENG based respiratory monitor with interdigital electrode was recently reported, presenting high durability and sensitivity. The further developed wearable monitor system provides real-time information covering respiratory rate, apnea, and respiratory ventilation.[14] With assistance of artificial intelligence, function advance of the TENG based system has been realized, including more parameters acquisition (such as vital capacity, peak expiratory flow) along with signal analysis and wireless data transfer.[15]

## 3. Potential application for wearable electronic products

Wearable electronics usually prefer good flexibility, recently all-fabric direct-current TENGs (AFDC-TENG) were fabricated with flexible semiconductor fabric made from cetyltrimethylammonium bromide and sodium dodecylbenzene sulfonate modified single-wall carbon nanotubes (SWCNTs).[16] AFDC-TENG presents high flexibility, comfort, and DC output voltage, current, and power density of 0.2 V, 0.29 μA, 45.5 mV m$^{-2}$.

## 4. Human-Machine Interface

The advances in artificial intelligence promote the research of human-machine interface, which serve as an important tool for human-machine interaction and communication. A TENG based voice and gesture signal translator (VGST), was developed recently by Wang's group, realizing a high sensitivity of 167 mV/dB and the resolution of 0.1 Hz by voice-electric signal conversion. With assistance of machine learning, the VGST presents accurate recognition performance for speech and simple hand gestures, indicating a promising prospect of TENG in the application of human-machine interface.[17]

## 5. Ambipolar transistors

TENGs involving in construction of ambipolar transistors was also reported recently. MoTe$_2$ FETs made through mechanical exfoliation with source, drain electrodes prepared via thermal evaporation deposition of Cr/Au at 1 or 2 Å·s$^{-1}$ is connected to TENG with Al-PTFE-Al structure at the gate electrode. The combination of MoTe$_2$ and TENG results in the transistor modulated by mechanical signal induced triboelectric potential with cutoff current as low as 1 pA·μm$^{-1}$ and current on/off ratio up to ~10$^3$ for both electron and hole transportation.[18]

## 6. Smart boxing bag

Elastic-material-based TENGs usually suffer from the low output and sensitivity, hindering its application in mechanical energy harvest. A gradient nano-doping strategy with polytetrafluoroethylene (PTFE) nanoparticles has been recently demonstrated to be effective to enhance charge density and sensitivity. With this strategy, charge density of 537 μC m$^{-3}$ and peak power density of 732.6 mW m$^{-3}$ was realized via a sponge TENG, which was further fabricated to smart boxing bag with real-time force analysis and counting functions.[19] The gradient nano-doping method can be expected to play a beneficial role in expanding the application fields of elastic-material-based TENGs.

## 7. Underwater Tactile Tensegrity

The advances of autonomous underwater vehicles (AUVs) have made it an indispensable necessity for underwater exploration. However, the fully perceiving unfamiliar underwater environments remains a challenge for existing AUVs due to the lack of effective tactile sensing. Interestingly, TENGs based underwater tactile tensegrity has recently been developed with deep-learning technology. Endowed with advantages of ultrahigh sensitivity, fast response and low cost, the tactile tensegrity presents effective detection of the real time information such as magnitude, orientation of water perturbations.[20] The proposed TENGs based tactile tensegrity paves an avenue for underwater exploration with tactile sensing performance.

## 8. Vibration Monitoring

Abnormal vibration of equipment and devices such as generators, transformers and transmission lines provides important health condition information. TENGs based vibration monitor systems with IR LED were recently developed, where the instantaneous output power via triboelectric layers charge release reaches up to 616 W and LED can be powered as a transmitter for wireless communication. Furthermore, smart phone can be integrated with the system enabling visual vibration monitoring on the screen.

## References

[1] Hu Y, Qiu H, Sun Q, et al. Wheel‐structured Triboelectric Nanogenerators with Hyperelastic Networking for High‐Performance Wave Energy Harvesting. Small Methods, 2023.

[2] Li X, Xu L, Lin P, et al. Three-dimensional chiral networks of triboelectric nanogenerators inspired by metamaterial's structure. Energy & Environmental Science, 2023, 16 (7): 3040-3052.

[3] Liang X, Liu S, Lin S, et al. Liquid‐Solid Triboelectric Nanogenerator Arrays Based on Dynamic Electric‐Double‐Layer for Harvesting Water Wave Energy. Advanced Energy Materials, 2023, 13 (24).

[4] Han J, Liu Y, Feng Y, et al. Achieving a Large Driving Force on Triboelectric Nanogenerator by Wave‐Driven Linkage Mechanism for Harvesting Blue Energy toward Marine Environment Monitoring. Advanced Energy Materials, 2022, 13 (5): 2203219.

[5] Su E, Li H, Zhang J, et al. Rationally Designed Anti‐Glare Panel Arrays as Highway Wind Energy Harvester. Advanced Functional Materials, 2023, 33 (17).

[6] Li H, Wen J, Ou Z, et al. Leaf‐Like TENGs for Harvesting Gentle Wind Energy at An Air Velocity as Low as 0.2 m s$^{-1}$. Advanced Functional Materials, 2023, 33 (11).

[7] Feng L, Wang Z L, Cao X, et al. Accordion-inspired parallelly assembled triboelectric nanogenerator: For efficient

biomechanical energy harvesting and music responding. Nano Today, 2023, 49: 101760.

[8]  Yao S, Wang S, Zheng M, et al. Implantable, Biodegradable, and Wireless Triboelectric Devices for Cancer Therapy Through Disrupting Microtubule and Actins Dynamics. Advanced Materials, 2023.

[9]  Li H, Chen C, Wang Z, et al. Triboelectric immunotherapy using electrostatic-breakdown induced direct-current. Materials Today, 2023, 64: 40-51.

[10] Shi X, Chen P, Han K, et al. A strong, biodegradable, and recyclable all-lignocellulose fabricated triboelectric nanogenerator for self-powered disposable medical monitoring. Journal of Materials Chemistry A, 2023, 11 (22): 11730-11739.

[11] Zu L, Wen J, Wang S, et al. Multiangle, self-powered sensor array for monitoring head impacts. Science Advances, 2023, 9 (20).

[12] Wei C, Cheng R, Ning C, et al. A Self‐Powered Body Motion Sensing Network Integrated with Multiple Triboelectric Fabrics for Biometric Gait Recognition and Auxiliary Rehabilitation Training. Advanced Functional Materials, 2023.

[13] Zhan T, Zou H, Zhang H, et al. Smart liquid-piston based triboelectric nanogenerator sensor for real-time monitoring of fluid status. Nano Energy, 2023, 111.

[14] Li C, Xu Z, Xu S, et al. Miniaturized retractable thin-film sensor for wearable multifunctional respiratory monitoring. Nano Research, 2023.

[15] Li Y, Liu C, Zou H, et al. Integrated wearable smart sensor system for real-time multi-parameter respiration health monitoring. Cell Reports Physical Science, 2023, 4 (1).

[16] Lv T, Cheng R, Wei C, et al. All-Fabric Direct-Current Triboelectric Nanogenerators Based on the Tribovoltaic Effect as Power Textiles. Advanced Energy Materials, 2023, 13 (29): 2301178.

[17] Luo H, Du J, Yang P, et al. Human-Machine Interaction via Dual Modes of Voice and Gesture Enabled by Triboelectric Nanogenerator and Machine Learning. Acs Applied Materials & Interfaces, 2023, 15 (13): 17009-17018.

[18] Li Y, Yu J, Wei Y, et al. Ambipolar tribotronic transistor of MoTe2. Nano Research, 2023.

[19] Gao X, Xing F, Guo F, et al. Strongly enhanced charge density via gradient nano-doping for high performance elastic-material-based triboelectric nanogenerators. Materials Today, 2023, 65: 26-36.

[20] Xu P, Zheng J, Liu J, et al. Deep-Learning-Assisted Underwater 3D Tactile Tensegrity. Research, 2023, 6.